\documentclass[conference]{IEEEtran}
\usepackage{cite}
\usepackage{graphicx}
\usepackage{amsmath}
\usepackage{array}
\usepackage{amssymb,hhline,enumerate,dsfont}

\newtheorem{theorem}{Theorem}

\begin{document}

\title{Optimum Linear LLR Calculation for Iterative Decoding on Fading Channels}

\author{\authorblockN{Raman Yazdani, Masoud Ardakani}
\authorblockA{Department of Electrical and Computer Engineering,
University of Alberta\\
Edmonton, Alberta T6G 2V4, Canada\\
Email: \{yazdani, ardakani\}@ece.ualberta.ca} }

\maketitle

\begin{abstract}
On a fading channel with no channel state information at the
receiver, calculating true log-likelihood ratios (LLR) is
complicated. Existing work assume that the power of the additive
noise is known and use the expected value of the fading gain in a
linear function of the channel output to find approximate LLRs. In
this work, we first assume that the power of the additive noise is
known and we find the optimum linear approximation of LLRs in the
sense of maximum achievable transmission rate on the channel. The
maximum achievable rate under this linear LLR calculation is almost
equal to the maximum achievable rate under true LLR calculation. We
also observe that this method appears to be the optimum in the sense
of bit error rate performance too. These results are then extended
to the case that the noise power is unknown at the receiver and a
performance almost identical to the case that the noise power is
perfectly known is obtained.
\end{abstract}

\section{Introduction}

Iterative decoding has received much attention in the past decade
due to its exemplary performance. There have been many advances in
iterative decoding techniques and it has been shown that using
graphical codes such as low-density parity-check (LDPC) codes
\cite{ldpc:gallager63} and turbo codes \cite{ldpc:berrou93}
associated with iterative decoding, the Shannon limit on many
channels (e.g., additive white Gaussian noise channel) can be
approached \cite{ldpc:chung01}. Therefore, these codes have also
been proposed for wireless fading channels \cite{ldpc:hou01}.

Application of LDPC codes on Rayleigh fading channel is pioneered in
\cite{ldpc:hou01}, where a detailed study of performance and code
design is conducted. This work is later extended to complex fading
channels \cite{ldpc:fu02}, to Rician fading channels
\cite{ldpc:lin06}, and also to Rayleigh block fading channels
\cite{ldpc:xiaowei04}. The application of turbo codes on Rayleigh
fading channels is also studied in \cite{ldpc:hall96}.

For soft iterative decoding, log-likelihood ratios (LLRs) at the
output of the channel are calculated. The process of computing LLRs
depends on whether or not a perfect knowledge of the channel
parameters exists at the receiver. The capacity of the fading
channel is also affected with the availability of channel parameters
at the receiver \cite{ldpc:hall96}.

An uncorrelated fading channel can be modeled with a fading gain $r$
and an additive Gaussian noise $n\sim\mathcal{N}(0, \sigma_n^2)$.
When $r$ is known at the receiver as a perfect side information
(SI), LLRs are linear functions of the channel output
\cite{ldpc:hou01}. Exact LLR computation depends on a perfect
knowledge of $\sigma_n$ at the receiver.

In order to have SI at the receiver, channel estimation techniques
must be used. These techniques increase the complexity of the
system, can cause a significant overhead, and are themselves subject
to imperfections. For high throughput wireless applications, the
receiver may not be able to handle the extra complexity or overhead.
This paper provides an alternative solution which does not require
channel estimation, yet provides better performance compared to the
existing solutions that use fixed fading gain estimates in the
decoder.

With no SI available at the receiver, LLRs are complicated functions
of the channel output \cite{ldpc:hagenauer80} and depend on the
probability density function (pdf) of $r$. An approximate LLR,
however, can be computed as a linear function of the channel output
\cite{ldpc:hagenauer80}. The coefficient of this linear function
depends on a fixed estimate $\hat r$ of the channel fading gain and
a knowledge of $\sigma_n$. Previous work assume that $\sigma_n$ is
known and use the expected value of $r$ for $\hat r$
\cite{ldpc:hou01,ldpc:hall96}. While the expected value of the
fading gain is the minimum mean square error estimation of $r$, it
is not guaranteed that this choice provides the optimum performance
in the decoder.

In a general setup (which includes famous fading channel models such
as uncorrelated Rayleigh and Rician fading channels), we propose the
following question: Assume that the pdf of $r$ is known at the
receiver, but the channel fading gain is not. Also assume that LLRs
are to be computed as linear functions of the channel output. What
linear approximation provides the optimum decoding performance?

This question is studied in this paper and the following
contributions are made: (1) When $\sigma_n$ is known, we find a
linear LLR approximation which allows for the maximum achievable
code rate on the channel. We prove that the optimum linear
approximation is unique and we observe that, on a Rayleigh fading
channel, it closely approaches the capacity under true LLR
calculation. This solution can significantly outperform LLR
calculation based on the expected value of $r$. We also design
irregular LDPC codes which approach this maximum achievable rate.
(2) When neither $r$ nor $\sigma_n$ is known at the receiver, we
propose a linear LLR calculation technique which guarantees the
convergence of the decoder over the widest possible range of
$\sigma_n$. The performance of this solution is almost identical to
the case that $\sigma_n$ is perfectly known. We design appropriate
irregular LDPC codes for this case too.


This paper is organized as follows. Section \ref{Sec:Preli} reviews
some preliminaries and studies the proposed approaches. Section
\ref{Sec:Known} studies the problem when $\sigma_n$ is known to the
receiver. Section \ref{Sec:Unknown} extends our results to the case
that $\sigma_n$ is unknown. Section \ref{Sec:Conclude} concludes the
paper.

\section{Preliminaries and Approaches} \label{Sec:Preli}

\subsection{System model}

Consider the following channel model. The output of the channel is
given by
\begin{equation}\label{channel_model}
    y=r\cdot x+n,
\end{equation}
where $x\in\{-1,1\}$ represents the input signal and $n$ is the
Gaussian noise with zero mean and variance $\sigma_{n}^2$. Also
$r\ge0$ is the channel gain which has an arbitrary pdf $p(r)$ and
changes independently from one channel use to another. Uncorrelated
fading channels fall into this system model where $r$ represents the
channel fading gain.

\subsection{LLR definition and distributions}
For soft decoding, LLRs are usually computed and used. Analysis of
some iterative decoders is based on the pdf of LLRs under the
assumption that the all-zero codeword ($x=+1$) is transmitted
\cite{ldpc:richardson01}.

For the model in (\ref{channel_model}), the conditional pdf of $y$
is given by

\begin{equation}\label{y_conditional_pdf}
    p(y|x,r)=\frac{1}{\sqrt{2\pi}\sigma_n}\exp{\left(-\frac{(y-x\cdot
    r)^2}{2\sigma_n^2}\right)},
\end{equation}
which represents a Gaussian distribution with mean $x\cdot r$ and
variance $\sigma_n^2$. We are interested to compute the channel LLRs
and their pdf.

\subsubsection{Ideal SI}

When we have ideal SI, the channel fading gain $r$ is known for each
received bit. Also, the receiver knows the noise power. Therefore,
the LLR is given by \cite{ldpc:hou01}
\begin{equation}\label{llr_si}
    l=\log\frac{P(x=+1|y,r)}{P(x=-1|y,r)}=\frac{2}{\sigma_n^2}y\cdot
    r,
\end{equation}
which is a linear function of $y$.

\subsubsection{No SI}

When no side information is available at the receiver, the channel
LLR is
\begin{equation}\label{llr_real_nsi}
    l=\log\frac{P(x=+1|y)}{P(x=-1|y)},
\end{equation}
which can be a complicated function of $y$ in general. For instance,
on a normalized Rayleigh channel (i.e., $p(r)=2re^{-r^2}$) we have
\begin{equation} \label{llr_real_nsi_Rayleigh}
l=\log\frac{\Phi(y/\sqrt{2\sigma_n^2(1+2\sigma_n^2)})}{\Phi(-y/\sqrt{2\sigma_n^2(1+2\sigma_n^2)})},
\end{equation}
where $\Phi(z)=1+\sqrt \pi ze^{z^2}\mathrm{erfc}(-z)$ and
$\mathrm{erfc}(\cdot)$ represents the complementary error function
\cite{ldpc:hagenauer80}. This LLR is a complicated function of $y$
and hard to be calculated in the decoder. Also, calculating the LLR
pdf is difficult. To simplify the LLR calculation, motivated by
(\ref{llr_si}), we write $\hat{l}$ as
\begin{equation}\label{llr_nsi}
    \hat{l}=\frac{2}{\hat{\sigma}_n^2}y\cdot \hat{r}=\alpha y,
\end{equation}
where $\hat{\sigma}_n^2$ represents the receiver's estimate of the
Gaussian noise variance $\sigma_n^2$ and $\hat{r}$ represents a
fixed receiver's estimate of the fading gain $r$. Here,
$\alpha=2\frac{\hat r}{\hat{\sigma}_n^2}$. This linear
representation of LLR is consistent with the results in
\cite{ldpc:hagenauer80} which states that the LLR can be
approximated by a linear function of $y$. This approach is also
consistent with existing work which assumes that $\sigma_n$ is known
and uses expected value of $r$ ($\mathrm{E}[r]$) as $\hat r$
\cite{ldpc:hou01}.

The conditional pdf of $\hat{l}$ is
\begin{equation}\label{llr_nsi_conditional_pdf}
    p(\hat{l}|r)=\frac{\hat{\sigma}_n^2}{2\hat{r}\sigma_n\sqrt{2\pi}}\exp
    \left(-\frac{(\hat{l}-2r\hat{r}/\hat{\sigma}_n^2)^2}{8\hat{r}^2\sigma_n^2/\hat{\sigma}_n^4}\right).
\end{equation}
To get the unconditional pdf of $\hat{l}$,
(\ref{llr_nsi_conditional_pdf}) should be averaged over the density
of $r$. For example, for the normalized Rayleigh fading channel we
have
\begin{eqnarray}\label{llr_nsi_pdf}
    \nonumber \lefteqn{p(\hat{l})_{\{\sigma_n,\alpha\}}=\frac{2\Delta^2}{\alpha\sigma_n\sqrt{2\pi}}\exp
    \left(-\frac{\Delta^2}{\alpha^2\sigma_n^2}\hat{l}^2\right)
    \times} \\
    &&\left[\exp\left(-\frac{\Delta^2}{2\alpha^2\sigma_n^4}\hat{l}^2\right)
    +\frac{\Delta\sqrt{2\pi}}{2\alpha\sigma_n^2}\hat{l}~\mathrm{erfc}\left(
    -\frac{\Delta}{\alpha\sigma_n^2\sqrt{2}}\hat{l}\right)\right]
\end{eqnarray}
where $\Delta=\sqrt{\frac{\sigma_n^2}{2\sigma_n^2+1}}$. This pdf is
parameterized by $\sigma_n$ and $\alpha$. If
$\hat{\sigma}_n=\sigma_n$ and $\hat{r}=\mathrm{E}[r]$, i.e.,
$\alpha=2\frac{\mathrm{E}[r]}{\sigma_n^2}$, (\ref{llr_nsi_pdf})
reduces to the distribution in \cite[Eq. 16]{ldpc:hou01}.

\subsection{Capacity}

The capacity of a binary-input memoryless symmetric channel (BMSC)
can be given via the pdf $p(l)$ of the LLR by \cite{ldpc:etesami06}
\begin{equation}\label{capacity}
    C=1-\mathrm{E}_l[\log_2{(1+e^{-l})}]=1-\int_{-\infty}^{\infty}\log_2{(1+e^{-l})}p(l)dl.
\end{equation}
The above relation is only valid for BMSCs where the LLR pdf is
consistent (i.e., $p(-l)=e^{-l}p(l)$). The channel capacity $C$ can
be computed in two cases: with ideal SI or no SI. In each case,
their corresponding LLR distribution should be used in
(\ref{capacity}). In the absence of SI at the receiver, the quantity
calculated by putting $p(\hat{l})$ in (\ref{capacity}) called
$\hat{C}$ is not the channel capacity since $\hat{l}$ is a linear
approximation and not the true LLR. Also, since $p(\hat{l})$ is not
consistent, $\hat{C}$ does not represent the highest achievable
transmission rate under linear LLR calculation of (\ref{llr_nsi}).
However, we observe that by maximizing $\hat{C}$ with respect to
$\alpha$, $p(\hat{l})$ nearly becomes a consistent distribution and
$\hat{C}$ predicts the maximum transmission rate under this optimum
linear LLR calculation quite accurately. This maximum $\hat{C}$ is
extremely close to $C$ in the absence of SI (see Fig.
\ref{fig:capacity}).

\subsection{LDPC codes decoding and analysis}
Some of the results of this paper are shown through analysis and
design of LDPC codes. Therefore, a quick review of LDPC codes seems
relevant. We use $\mathcal{C}^N(\lambda(x),\rho(x))$ to denote an
ensemble of LDPC codes of length $N$ with variable and check node
degree distributions $\lambda(x)$ and $\rho(x)$ respectively
\cite{ldpc:richardson01}.

Many different message-passing algorithms can be used for the
decoding of LDPC codes. In this work, our focus will be on the
\emph{sum-product} algorithm \cite{ldpc:kschischang01}.

For the channel model of (\ref{channel_model}), the \emph{decoding
threshold} $\sigma_n^\ast$ of an ensemble of LDPC codes is defined
as the maximum noise standard deviation $\sigma_{n}$ for which the
bit error probability of the message-passing decoder gets
arbitrarily small when the code length is growing
\cite{ldpc:richardson01_2,ldpc:richardson01} if and only if
$\sigma_n\le\sigma_n^\ast$. This $\sigma_{n}^\ast$ depends on
whether SI is available at the receiver or not.

The most exact LDPC code analysis is density evolution, which takes
the pdf of the channel LLRs and tracks the evolution of the pdf of
the decoder's extrinsic messages in each iteration
\cite{ldpc:richardson01_2,ldpc:richardson01}. Formulation of this
method in closed form is too complex, hence, some numerical
approximations are often used \cite{ldpc:chung01}.

\subsection{Code design}
It is well known that carefully designed irregular LDPC codes can
approach the capacity of many channel models (e.g., see
\cite{ldpc:chung01}). Two code design processes associated with two
measures of performance can be defined one as maximizing the
threshold of the code over its degree distributions given a target
code rate and another one, as maximizing the code rate over its
degree distributions given the channel LLR pdf.

\section{Optimum linear LLR calculation} \label{Sec:Known}

As mentioned before, when no SI is available at the receiver, one
can calculate the LLRs linearly via (\ref{llr_nsi}) as an
approximation to (\ref{llr_real_nsi}). The objective is to find the
optimal linear approximation. Different measures of optimality can
be considered. Existing work assumes that $\sigma_n$ is known and
chooses $\hat{r}=\mathrm{E}[r]$. This choice of $\hat r$ is optimum
in the sense of minimum mean square error
$\mathrm{E}[|r-\hat{r}|^2]$. In this work, we find the linear
approximation which gives a nearly consistent LLR pdf and results in
the maximum achievable transmission rate on the channel. We call
this linear approximation \emph{maximum-capacity}
linear-approximation (MCLA).

Different linear approximations result in different LLR
distributions for $\hat l$. Each LLR distribution defines a
corresponding $\hat C$ according to (\ref{capacity}). Thus, the
problem of finding the MCLA is simplified to finding a linear
approximation whose corresponding LLR distribution maximizes the
capacity $\hat C$.

Maximizing $\hat C$ requires a knowledge of $\sigma_n$ and pdf of
$r$. These are needed for finding the pdf of $\hat l$ and thus
optimizing its corresponding capacity. So, we first assume that
these pieces of information are available. Later we generalize our
results to the case that $\sigma_n$ is unknown. When $\sigma_n$ is
known, without loss of generality we set $\hat{\sigma}_n = \sigma_n$
in (\ref{llr_nsi}) and we find the optimum choice of $\hat{r}$.
Notice that with  $\hat r$ one can adjust $\alpha$ and thus the pdf
of $\hat l$ as needed.

MCLA maximizes a bound on the achievable transmission rate. To show
that this optimization is meaningful in practice, we design
irregular LDPC codes that approach this maximized capacity. More
interestingly, we observe that the optimized $\hat C$ is extremely
close to $C$ based on true LLR calculation.

MCLA is also expected to result in improved performance in iterative
decoders. That is, for a fixed code, computing LLRs according to
MCLA should improve BER performance. Our simulation results will
support this claim, but the following two arguments can also be
provided to justify this choice. (1) MCLA provides the maximum $\hat
C$ and thus the maximum gap between the code rate $R$ and the
capacity $\hat C$. Thus one expects improved BER performance. (2)
Since $\hat l$ is not the true LLR, under any linear LLR
calculation, $\hat C \le C$. Under a good linear approximation, pdf
of $\hat l$ is close to that of true LLRs and thus $\hat C$ is close
to $C$. Hence, a minimized $C - \hat C$ (through maximizing $\hat
C$) indicates a good LLR approximation.

As mentioned, our simulation results show that MCLA indeed improves
the performance compared to existing work based on choosing
$\hat{r}=\mathrm{E}[r]$. Moreover, though not rigorously proved,
MCLA appears to be the optimum choice in terms of BER performance
too. Thus, our proposed method is based on maximizing $\hat C$ over
$\hat r$ for fixed $\hat{\sigma}_n$ and $\sigma_n$, and we define
\begin{equation}\label{r_opt}
    \hat{r}_\mathsf{opt}=\arg \max_{\hat{r}}\hat{C}.
\end{equation}

The following theorem suggests that finding $\hat{r}_\mathsf{opt}$
can be done very efficiently.
\begin{theorem}
For a fixed $\hat{\sigma}_n$ and $\sigma_n$, there exists a unique
$\hat{r}$ which maximizes
$\Hat{C}=1-\mathrm{E}_{\hat{l}}[\log_2{(1+e^{-\hat{l}})}]$.
\end{theorem}
\emph{Proof:}
\begin{equation*}
    \Hat{C}=1-\mathrm{E}_{\hat{l}}[\log_2{(1+e^{-\hat{l}})}]=1-\mathrm{E}_y[\log_2{(1+e^{-\frac{2\hat{r}}{\hat{\sigma}_n^2}y})}]
\end{equation*}
\begin{eqnarray*}
    \frac{d^2\Hat{C}}{d{\hat{r}}^2} &=& -\mathrm{E}_y\left[\frac{d^2}{d{\hat{r}}^2}
    \left(\log_2{(1+e^{-\frac{2\hat{r}}{\hat{\sigma}_n^2}y})}\right)\right]
    \\
    &=& \mathrm{E}_y\left[\frac{-\left(\frac{2y}{\hat{\sigma}_n^2}\right)^2
    e^{-\frac{2\hat{r}}{\hat{\sigma}_n^2}y}}{\left(1+
    e^{-\frac{2\hat{r}}{\hat{\sigma}_n^2}y}\right)^2\ln2}\right]~<~0
\end{eqnarray*}

The above expression is negative since the term inside the expected
value is always negative. Therefore, $\hat{C}$ is a concave function
of $\hat{r}$ and there exists a unique maximum in
$\hat{r}=\hat{r}_\mathsf{opt}$. This theorem is valid for any
distribution of $r \ge 0$. \hspace{\stretch{1}}$\blacksquare$

\smallskip

\begin{figure}
  \centering
  \includegraphics[width=.99\columnwidth]{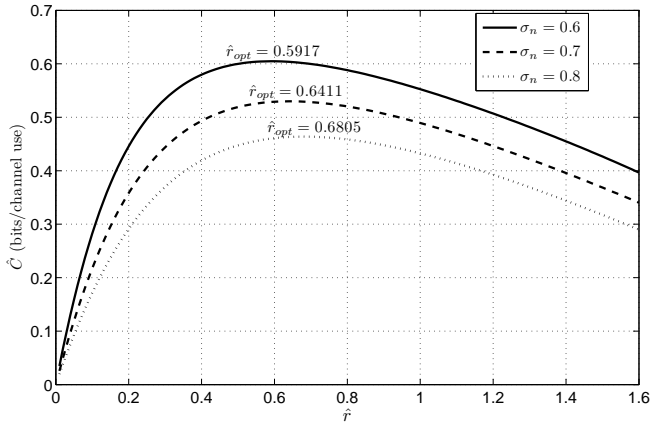}\\
  \caption{$\hat{C}$ for different $\sigma_n$ and $\hat{r}$ on a normalized Rayleigh fading channel. The curves are concave and the maximizing point is unique. Moreover,
the maximizing point is not much
  sensitive to $\sigma_n$.}\label{fig:capacity_r}
\end{figure}

Maximizing $\hat{C}$, therefore, is a straightforward task because
it is a one-variable convex-optimization problem and can be solved
very efficiently by simple numerical techniques. Different $\hat{C}$
curves are depicted in Fig. \ref{fig:capacity_r} for some $\hat{r}$
and the case $\hat{\sigma}_n=\sigma_n$. Notice that
$\hat{r}_\mathsf{opt}$ is not very sensitive to $\sigma_n$.

\begin{figure}
  \centering
  \includegraphics[width=.99\columnwidth]{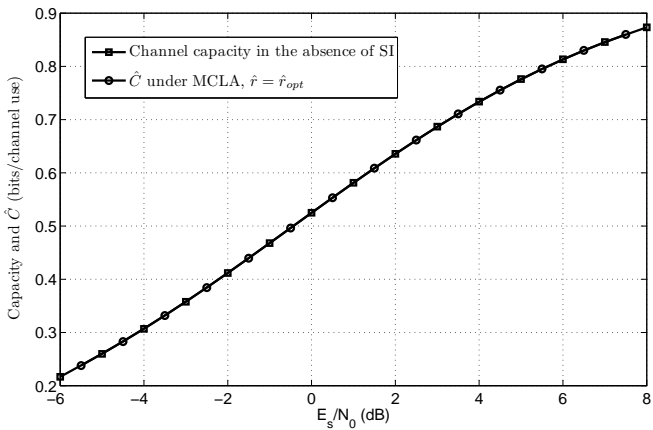}\\
  \caption{Comparison between the highest achievable transmission rate in the case of true LLR
  calculation and $\hat{C}$ on a normalized Rayleigh fading channel. It is assumed that $\hat{\sigma}_n=\sigma_n$.}\label{fig:capacity}
\end{figure}

Fig. \ref{fig:capacity} shows that we can get very close to the
channel capacity under MCLA. Simulations show that
$\hat{r}=\mathrm{E}[r]$ can result in significant performance loss
especially when the capacity or the code rate increases. This
performance loss is about 0.24~dB in 0.5~bits/channel use to 0.92~dB
in 0.75~bits/channel use.

The following two examples support our above mentioned results. In
Example 1, we design an irregular LDPC code which approaches the
capacity that is maximized by MCLA. Example 2 shows improved BER
performance under MCLA.

\emph{Example 1:} Consider an uncorrelated normalized Rayleigh
fading channel with $\sigma_n=0.7436$. The channel capacity is
0.5~bits/channel use in the absence of SI and $\hat{C}=0.4999$ using
$\hat{r}_\mathsf{opt}=0.6594$ (compare with $\mathrm{E}[r]=0.8862$).
We design a code based on rate maximization under MCLA. We assume a
fixed $\rho(x)=x^8$ and a maximum variable node degree $d_v$ of 30.
Under MCLA and 11-bit decoding, and allowing a maximum of 300
iterations, the optimized code is given in Table \ref{Table}
(Code1). The rate of the designed code is 0.4889. The designed code
has almost approached $\hat{C}$ and also the capacity of the channel
with no SI.

\begin{figure}
  \centering
  \includegraphics[width=0.99\columnwidth]{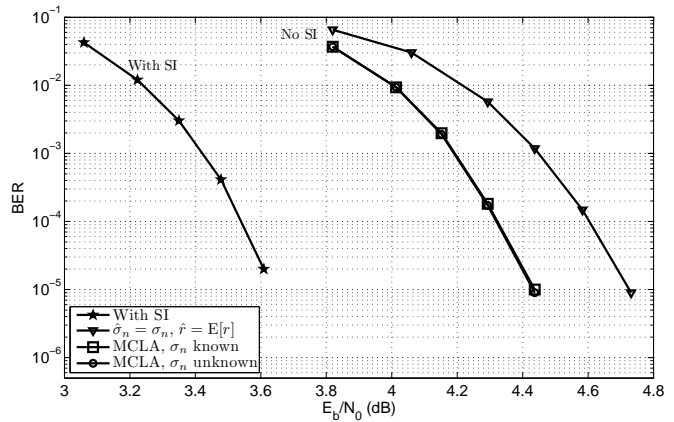}\\
  \caption{Comparison of BER for a $\mathcal{C}^{10000}(x^2,x^5)$ LDPC code in different cases on a normalized
  Rayleigh fading channel. The performance of MCLA remains almost
  the same regardless of whether $\sigma_n$ is known or not.
  }\label{fig:BER}
\end{figure}

\emph{Example 2:} To show that MCLA also improves the BER of the
code, a $\mathcal{C}^{10000}(x^2,x^5)$ LDPC code is simulated on an
uncorrelated normalized Rayleigh fading channel. Fig. \ref{fig:BER}
shows the BER of the code with and without SI at the receiver. When
SI is not available and $\sigma_n$ is known, two cases have been
plotted. One is when $\hat{\sigma}_n=\sigma_n$ and
$\hat{r}=\mathrm{E}[r]$ and the other is under MCLA (i.e.,
$\hat{\sigma}_n=\sigma_n$ and $\hat{r}=\hat{r}_\mathsf{opt}$). The
decoding threshold of the code is 4.06~dB with
$\hat{r}=\mathrm{E}[r]$ and no SI, 3.82~dB under MCLA, and 3.06~dB
with perfect SI. The figure shows considerable BER improvement under
MCLA. We have chosen a (3,6)-regular LDPC code since most of its
results exist in the literature. If a higher rate code (e.g.,
$R=3/4$) is used, the performance improvement increases.

\section{Noise power unknown at the receiver} \label{Sec:Unknown}
Under MCLA, the pdf of $\hat l$ is a linear transformation of the
pdf of $y$. In fact,
$\alpha_\mathsf{opt}=2\frac{\hat{r}_\mathsf{opt}}{\sigma_n^2}$ would
give the linear transformation whose associated capacity (given by
(\ref{capacity})) is maximum. When $\sigma_n$ is unknown, the
distribution of $y$ is not known at the receiver. Therefore, finding
the optimum linear transform is somewhat meaningless. However, for
any $\sigma_n$ one can find $\alpha_\mathsf{opt}$. Thus,
$\alpha_\mathsf{opt}$ is a function of $\sigma_n$. It is also
obvious from Theorem 1, that $\alpha_\mathsf{opt}$ is unique for
each $\sigma_n$. Thus, $\alpha_\mathsf{opt}$ will be denoted as
$\alpha_\mathsf{opt}(\sigma_n)$ afterwards.

Since $\sigma_n$ is unknown,  $\alpha_\mathsf{opt}(\sigma_n)$ is
also unknown. However, one can find $\alpha$ such that a given code
has the widest range of convergence over changes of $\sigma_n$. This
way, we ensure that the code is robust to the changes in the noise
power (e.g. when the code is used in different channels with
different noise powers).

The basic idea for finding such $\alpha$ is to maximize the
achievable transmission rate at the highest noise standard deviation
that the code can tolerate under MCLA. To do this, for a given code,
we must find the largest $\sigma_n$, referred to as $\sigma_n^\ast$,
such that the code still converges to zero error rate when LLRs are
obtained using (\ref{llr_nsi}) with
$\alpha=\alpha_\mathsf{opt}(\sigma_n^\ast)$. Finding $\sigma_n^\ast$
can be done efficiently through a binary search, but
at each stage of the search $\alpha_\mathsf{opt}$ must be updated
accordingly.

This choice of $\alpha$ gives the widest convergence range over
$\sigma_n$, because it is the optimum $\alpha$ in the worst channel
condition. When the channel condition improves, this choice of
$\alpha$ is no longer optimum. But we expect that even with a
sub-optimal $\alpha$, convergence is achieved due to improvement in
the channel condition. This can also be justified recalling that
$\hat{r}_\mathsf{opt}$ (and hence $\alpha_\mathsf{opt}$) is not very
sensitive to $\sigma_n$. Our simulation results on LDPC codes will
confirm that this choice of $\alpha$ provides the widest convergence
range.

In order to measure the performance we do as follows. For various
$\sigma_n$ and different values of $\alpha$ (including
$\alpha_\mathsf{opt}(\sigma_n)$), we find the required number of
density evolution iterations to achieve a target message error-rate
(MER) $p_t$ for a given LDPC code. We use the required number of
iterations $\ell^{\ast}(p_{t})$ as a comparison measure and to
identify the range of convergence.

\begin{figure}
  \centering
  \includegraphics[width=.99\columnwidth]{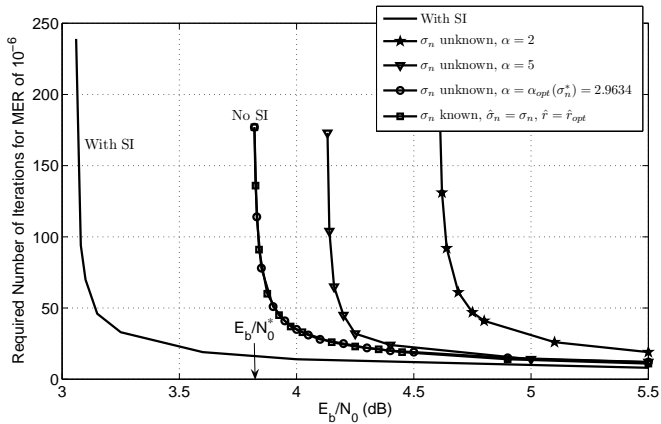}\\
  \caption{Comparison between the performances of $\mathcal{C}^\infty(x^2,x^5)$
  on a normalized Rayleigh fading channel with or without SI using different $\alpha$.}\label{fig:de_3_6}
\end{figure}

In Fig. \ref{fig:de_3_6}, different values of $\alpha$ are used and
$\ell^{\ast}(p_{t})$ is plotted for different values of $\sigma_n$
for $\mathcal{C}^\infty(x^2,x^5)$. It is seen that by using
$\alpha=\alpha_\mathsf{opt}(\sigma_n^\ast=0.6442)=2.9634$, the code
has the widest convergence range. Interestingly, while this choice
of $\alpha$ is not optimum for all values of $\sigma_n$, the
resulted $\ell^{\ast}(p_{t})$ is always very close to the curve
based on known $\sigma_n$ under MCLA. This observation can also be
made from Fig.~\ref{fig:BER}. The threshold of this code is at
3.06~dB with SI and is at 3.82~dB with no SI under MCLA. Thus, the
gap between these thresholds is 0.76~dB. At this code rate,
existence of SI results in about 0.74~dB improvement
\cite{ldpc:hall96, ldpc:hou01}. Thus, MCLA shows a minor extra gap
(0.02~dB) compared to true LLR calculation.

When $\sigma_n$ is unknown, one can design an LDPC code with a given
rate which under MCLA provides the widest convergence region, i.e.,
has the largest decoding threshold. This procedure has to be done
carefully, because $\alpha_\mathsf{opt}$ is a function of
$\sigma_n^\ast$ which is initially unknown. We omit the details of
this code design procedure in the interest of available space, but
one designed code is reported in Table \ref{Table} (Code2). Again,
11-bit decoding under MCLA is used, the maximum number of iterations
allowed is 300 and $d_v=30$. The threshold of the designed code is
2.76~dB. This code has the largest decoding threshold among all the
codes with the rate 0.5.

\begin{table}[t]
\begin{center}
\begin{tabular}{|c|c|c|}
  \hline
                & Code1 & Code2 \\ \hline
  $\lambda_2 ,\lambda_3$  & 0.1916, 0.2244 & 0.1943, 0.2341 \\
    $\lambda_4,\lambda_5$ & 0.0057, 0.0109 & 0.0064, 0.0113 \\
    $\lambda_6, \lambda_7$ & 0.0427, 0.1187 & 0.0340, 0.0994 \\
    $\lambda_8, \lambda_9$ & 0.0297, 0.0121 & 0.0474, 0.0205 \\
    $\lambda_{10}, \lambda_{11}$ & 0.0147, 0.0000 & 0.0119, 0.0171 \\
    $\lambda_{15}, \lambda_{20}$ & 0.0157, 0.0314 & 0.0228, 0.0627 \\
    $\lambda_{29}, \lambda_{30}$ & 0.0382, 0.2649 & 0.0680, 0.1715 \\ \hline
    $\rho_9$ & 1.0000 & 1.0000 \\ \hline
    Rate & 0.4889 & 0.5000 \\
    $\sigma_n^\ast$ & 0.7436 & 0.7274  \\
  ${E_b/N_0}^\ast$ (dB) & 2.57 & 2.76 \\
  \hline
\end{tabular}
\caption{Good LDPC codes designed under MCLA. Code1 is a rate
maximized and Code2 is a threshold maximized code.} \label{Table}
\end{center}
\end{table}

\section{Conclusion} \label{Sec:Conclude}
We proposed a new method for linear LLR calculation on fading
channels when channel fading gain is not known at the receiver. Our
method is optimum in the sense of maximum achievable rate on the
channel. We showed that on a Rayleigh channel, the maximum
achievable rate using this method is extremely close to the channel
capacity. Compared to existing work, which uses the expected value
of the fading gain for LLR calculation, we reported considerable
performance improvement at no extra decoding cost. This improvement
would become more significant when the code rate increases.


We then extended our approach to the cases that the additive noise
power of the fading channel is also unknown at the receiver. With a
careful choice of linear LLR calculation, we were able to obtain a
performance almost identical to the previous case, where the
additive noise power was known.

For applications that channel estimation results in significant
overheads or suffers from severe imperfections, our proposed
solution can be of interest.

While we verified some of our results through study and design of
LDPC codes on Rayleigh channel, our approach for maximizing the
achievable transmission rate and convergence range of the decoder is
general. The only reason for using LDPC codes is that, they can
approach theoretical limits and thus verify some of our asymptotic
results.

\bibliographystyle{IEEEtran}
\bibliography{IEEEabrv,ldpc}

\end{document}